\documentclass[prb,twocolumn,aps,nobibnotes,showpacs]{revtex4}
\usepackage[final]{graphicx}

\newcommand{\cdti}{CdTiO$_3$}
\newcommand{\pbnm}{$Pbnm$}
\newcommand{\pbnv}{$Pbn2_1$}
\newcommand{\cgav}{V$_{33}$}
\newcommand{\unitefg}{\mbox{10$^{21}$V/m$^2$}}
\begin{document}
\title{First-principles study of orthorhombic CdTiO$_3$ perovskite}

\author{G. Fabricius}
\email{fabriciu@venus.fisica.unlp.edu.ar} 
\author{A. L\'opez Garc\'{\i}a}
\email{abeti@venus.fisica.unlp.edu.ar} 
\affiliation{
Depto. de F\'{\i}sica, Univ. Nacional de La Plata,\\
1900 La Plata, Argentina
}
\date{\today}

\begin{abstract}
In this work we perform an
{\it ab-initio} study of \cdti\ perovskite in its orthorhombic
phase using FLAPW method. Our calculations help to decide between
the different cristallographic structures proposed for this perovskite
from X-Ray measurements. 
We compute the electric field
gradient tensor (EFG) at Cd site and obtain excellent agreement with 
available experimental information from a perturbed
angular correlation (PAC) experiment. We study EFG under an isotropic
change of volume and show that in this case
the widely used "point charge model approximation" to determine
EFG works quite well. 
\end{abstract}


\maketitle 

Perovskites materials of the form ABO$_3$ have been intensively 
studied last years. In addition to the interest that 
present the technological aplications of ferroelectric materials,
perovskites are known to undergo several phase transitions
as a function of temperature
that make them appealing for theoreticians and experimentalists.
 The transitions between different structures usually
      involve very little atomic displacements
and are difficult to characterize, requiring some times 
several experimental techniques. First principles calculations
have been playing an increasing role in the understanding of the
these systems \cite{reviewvander}. 
These studies not only try to give theoretical 
explanations to the observed phenomena, but usually give valuable
information that help in the interpretation of the 
experimental results.

   \cdti\ perovskite has been the object of several studies 
   using X-Ray diffraction \cite{megaw,kaymiles,sasaki}, 
   dielectric properties methods \cite{martin,Sholokhovich,sasaki},
   infrared spectroscopy \cite{knyazev} and 
   PAC technique   \cite{baumvol,catchen}
   and there have been some
   difficulties and controversies in determining the symmetry group of it's 
   orthorhombic structure at room temperature.
      In 1957 H.F.Kay and J.L.Miles established through X-Ray
   measurements that \cdti\ perovskite should have orthorhombic
   symmetry and non-centrosymmetric $Pbn2_1$ \cite{pbn21} space group 
   with a large displacement of Ti atoms along the $b$ axis \cite{kaymiles}.
   However, the measurement of the electrical properties
   of \cdti\ single crystals
   couldn't resolve whether they were ferroelectric or not, then,
   the displacement of Ti atoms from their symmetric positions
   couldn't be confirmed \cite{Sholokhovich}. 
   In Reference\ \onlinecite{baumvol} Baumvol and coworkers study the EFG
   at Cd in \cdti\ through PAC measurements. Using the Point Charge
   Model approximation (PCM) to estimate theoretically the EFG
   they infer that the structure of
   the perovskite couldn't be the one corresponding to 
   \pbnv\  space group but should be another one
   with centrosymmetric space group, also considered and discarded by
   Kay\&Miles. Baumvol {\it et al.} made their arguments
   based on the asymmetry of EFG tensor, but in fact they obtained a 
   disagreement of about 80\% in the absolute value of EFG components.
     In 1987 S.Sasaki {\it et al.} \cite{sasaki} performed new X-Ray
     studies on \cdti .
    They obtained that 
   two orthorhombic structures with very similar coordinates and different
   space groups (centrosymmetric \pbnm\ and non-centrosymmetric \pbnv)
   fitted equally well the X-ray data. Analizing some structural and
   physical properties, they finally argued that it was more suitable to
   assign the space group \pbnm\ to \cdti.

   The main objective of the present work is to clarify
   this picture, studying from
   first-principles the electronic properties of the
    different
   structures proposed for the \cdti\ perovskite.

      Other important question that may be learned from this system
   concerns the correctness of the predictions made by FLAPW 
   and PCM approximation to the EFG tensor at Cd.
   With PAC technique the EFG tensor at a probe introduced in the solid
   can be determined \cite{pac}, and
   Cd is one of the most used PAC probes, so, it is of valuable 
   interest to check the different theoretical
   approaches to the problem of determining EFG at a Cd site.
   In the PCM the EFG at the probe  is usually approached as:
\begin{equation}
V_{ij}=(1 - \gamma _{\infty}) V_{ij}^{latt.}
\end{equation}
where $V_{ij}^{latt.}$ is the EFG produced by the valence charges 
located as point charges at the ions positions. 
In this approximation the only effect
of the probe is to antishield the lattice contribution through the
Sternheimer antishielding factor \cite{sternheimer}, that for Cd is usually taken as:
$\gamma _{\infty}=-29.27$ \cite{feiock}.
       PCM has been extensively 
   used to predict EFG at Cd sites in perovskites
    and oxides \cite{forker,wiarda,delafossites,lacro3} but
    calculations at the {\it ab-initio} level are rare in 
    these systems because Cd usually enters the solid as 
    impurity and this introduces several complications, as
    the effect of atomic relaxations, for example \cite{errico}. 
    In the case of \cdti\ perovskite, Cd 
  is one of the constituents of the system, so,  accurate
  {\it ab-initio} calculations can be performed and used to check
  the validity of PCM aproximation.

The first-principles calculations were performed
with the WIEN97 implementation of the Full-Potential
Linearized-Augmented-Plane-Wave (FLAPW) method
developed by
Blaha et al. \cite{wien97}. 
We use the Generalized Gradient Approximation (GGA) to 
describe the exchange correlation potential in the
parametrization of Perdew {\it et al.} \cite{gga}.
The size of the basis set in these calculations is 
controlled by the parameter $RK_{MAX}$ that we 
fixed in the value of 8. 
 The atomic sphere
  radii where chosen as 1.06, 0.90 and 0.85 \AA\
  for Cd, Ti and O atoms respectively. 
Integrations in reciprocal space were performed 
with the tetrahedron method using 18 and 36 $k$-points 
in the irreducible first Brillouin zone for
\pbnm\ and \pbnv\ structures respectively. 
  The unit cells have 20 atoms containing 5 molecules
  of \cdti. In the case of 
\pbnm\ structure there are 4 inequivalent atoms in 
the cell:
Cd($x_{\rm Cd}$,$y_{\rm Cd}$,$\frac{1}{4}$), 
Ti($\frac{1}{2}$,0,0), O1($x_{\rm O1}$,$y_{\rm O1}$,$\frac{1}{4}$) and 
O2($x_{\rm O2}$,$y_{\rm O2}$,$z_{\rm O2}$), while
for \pbnv\ structure there are 5 inequivalent atoms:
Cd($x_{\rm Cd}$,$y_{\rm Cd}$,$\frac{1}{4}$), 
Ti($x_{\rm Ti}$,$y_{\rm Ti}$,$z_{\rm Ti}$),
O1($x_{\rm O1}$,$y_{\rm O1}$,$z_{\rm O1}$), 
O2($x_{\rm O2}$,$y_{\rm O2}$,$z_{\rm O2}$) and            
O3($x_{\rm O3}$,$y_{\rm O3}$,$z_{\rm O3}$). 
Once self-consistency of the potential is achieved
the $V_{ij}$ componentes of the EFG tensor are 
computed from the lattice
harmonic expansion of the self-consistent potential \cite{efg}.
The EFG tensor is then diagonalized and it is expressed
in terms of its greatest absolute value component: $V_{33}$ and
the asymmetry parameter: $\eta=(V_{22}-V_{11})/V_{33}$, where
$V_{ii}$ are the components of the diagonalized tensor
that are ordered such that: $|V_{33}|>|V_{22}|>|V_{11}|$.

\begin{table*} 
\caption{\label{estruc}
Absolute value of forces 
on inequivalent atoms (F) in $eV/$ \AA, component
of largest absolute value of the EFG tensor (\cgav) in \unitefg\
and asymmetry parameter
($\eta$) calculated in the present work compared with PAC experiment.}
\begin{ruledtabular}
\begin{tabular}{lccccccc}
       &  F(Cd) & F(Ti) & F(O1) & F(O2) & F(O3) & \cgav & $\eta$ \\
			\hline 			  
\pbnv: Ref.~\onlinecite{kaymiles} &
         1.84  & 3.43  & 3.85  & 1.04  & 3.63  & +10.00  & 0.80   \\ 
\pbnm: Ref.~\onlinecite{sasaki} &
         0.03  & 0.00  & 0.05  & 0.05  &       &   -5.10 & 0.42   \\ 
\pbnv: Ref.~\onlinecite{sasaki} &
         0.03  & 0.10  & 0.41  & 0.45  & 0.54  &  -5.14  & 0.42   \\ 
PAC exp.\cite{baumvol} &
               &       &       &       &       &   5.41$\pm$0.80 & 0.407$\pm$0.008    \\ 
\end{tabular}
\end{ruledtabular}
\end{table*}

In Table~\ref{estruc} we show the results of our FLAPW calculations 
performed for the different structures reported in the literature compared
with the PAC determination of $|V_{33}|$ and $\eta$. 
We see that the huge value obtained for the forces in the case of
\pbnv\ structure of Ref.~\onlinecite{kaymiles}  and the poor agreement 
of $eq$ and $\eta$ with experiment lead us to discard this structure
as the stable one for \cdti\ perovskite. In the case of structures of 
Ref.~\onlinecite{sasaki}, we see that for both of them we obtained 
 $eq$ and $\eta$ in excellent agreement with the experimental result
 of Baumvol {\it et al.} \cite{baumvol}. The forces obtained for these
 structures are much smaller 
 than those obtained for \pbnv\ structure of Ref.~\onlinecite{kaymiles}  
 but the ones on oxygen atoms for \pbnv\ structure of Ref.~\onlinecite{sasaki}
 are still too big to be attributed to the precision 
 of our calculations. So, in order
 to check the stability of structures of Ref.~\onlinecite{sasaki} 
 in the frame of FLAPW calculations, we have relaxed both structures 
 (keeping fixed the lattice parameters at their experimental values) until
 forces on every atom are below a tolerance value taken as $0.01 eV/$ \AA.
 These relaxation processes involve the variation of the 7 and 14 independent
coordinates that define \pbnm\ and \pbnv\ structures respectively, and we
 have performed them following a Newton damped scheme \cite{newton}. 
 At the end of the relaxations we obtained that both structures
 relaxed to the same \pbnm\ structure within a tolerance in distances
 of $0.01$ \AA. The finally obtained \pbnm\ structure 
was very similar to the one proposed by Sasaki {\it et al.}, as expected,
due to the small forces previously obtained for this 
structure shown in Table~\ref{estruc}. 
So, our calculations indicate that \pbnv\ structure is not stable and confirm 
the assignation of \pbnm\ as the space group of
\cdti\ perovskite that was suggested in Ref.~\onlinecite{sasaki}.
The fact that X-Ray data from 
Ref. ~\onlinecite{sasaki} are also adjusted quite well by a non-centrosymmetric
structure with oxygens O2 and O3 that slightly appart
from their corresponding symmetric positions
in \pbnm\ structure may indicate
the presence of a soft mode in \pbnm\ structure 
involving those displacements. A
future study of the phonons in the \pbnm\ structure would be
valuable in order to check this hypothesis.

\begin{table*} 
\caption{\label{efgs}
Electric field gradients as a function of an isotropic
change of volume for the \pbnm\ structure of \cdti\ perovskite.
PCM calculations were performed on the FLAPW-relaxed structures.
$V/V_0$ is the relation between volumes taken in our
calculations and those of Ref.~\onlinecite{sasaki}, $d_{nn}(Cd-O)$ (in \AA)
is the nearest-neighbour Cd-O distance.
\cgav\ and $\eta$, as in Table~\ref{estruc}.} 
\begin{ruledtabular}
\begin{tabular}{lccccccc}
 & & \multicolumn{3}{c}{FLAPW} & \multicolumn{3}{c}{PCM} \\  
$V/V_0$ & $d_{nn}(Cd-O)$ & \cgav\ & $\eta$ & \cgav -direction 
                         & \cgav\ & $\eta$ & \cgav -direction \\
			\hline 			  
0.913 & 2.18  & -5.63  &  0.49  &  (0.759, 0.651, 0.) & -4.67 & 0.59  & (0.773, 0.634, 0.) \\
0.956 & 2.21  & -5.33  &  0.44  &  (0.772, 0.636, 0.) & -4.53 & 0.52  & (0.778, 0.629, 0.) \\
1.000 & 2.25  & -5.04  &  0.40  &  (0.787, 0.617, 0.) & -4.35 & 0.47  & (0.784, 0.620, 0.) \\
1.046 & 2.28  & -4.82  &  0.40  &  (0.807, 0.591, 0.) & -4.21 & 0.45  & (0.795, 0.607, 0.) \\
1.093 & 2.31  & -4.62  &  0.38  &  (0.829, 0.560, 0.) & -4.07 & 0.41  & (0.809, 0.587, 0.) \\
\end{tabular}
\end{ruledtabular}
\end{table*}

As was mentioned, Baumvol {\it et al.} 
analizing the asymmetry of the EFG tensor with the
point charge model
supposed \cdti\ should have 
a centrosymmetric space group,
but in fact
they obtained \cgav\ larger than experiment, at least, in 80 \%.
In effect, the discrepancy in \cgav\ would be {\em only} about 80 \%
assuming the same sign for PCM prediction and experimental value,
and they atributted this discrepancy to a failure of PCM to 
account for a suppossed existing covalent contribution to EFG
at Cd in this system. But the sign of \cgav\ predicted by
Baumvol is opposite to the one that emerges from FLAPW calculations.
The point is that they used coordinates of a centrosymmetric structure
studied in Ref.~\onlinecite{kaymiles}.
If the coordinates obtained later by Sasaki {\it et al.}
\cite{sasaki} are used in the PCM calculations, 
\mbox{\cgav =-4.41} and $\eta$=0.45
are obtained in fairly good agreement with experiment and FLAPW calculations.
In order to inspect further this agreement between PCM and FLAPW predictions
to EFG tensor, we perform FLAPW calculations for an 
isotropic change of volume of the \pbnm\ structure.
 For each volume considered we relaxed
the internal coordinates and evaluate EFG tensor 
at the relaxed coordinates. The change is isotropic in the sense that
we hold the relations $a/b$ and $a/c$ fixed in the experimental 
ratios of Ref.~\onlinecite{sasaki}. The results are shown in Table~\ref{efgs}.
Both EFG estimations predict an increase of \cgav\ and $\eta$ with compression
and also an increase of the angle between \cgav -direction and x-axis. 
PCM not only predicts the same tendence 
as FLAPW, but also the absolute values are in fairly
good agreement. However, the agreement of \cgav\ and especially $\eta$ 
seems to be better for larger Cd-O distances where covalency between
Cd and oxygen is smaller.
In Ref.~\onlinecite{wiarda}, using PCM to compare with several
PAC measurements in oxides, 
the authors obtained empirically
 that when
Cd-O bond is larger than 2.1 \AA\ a good agreement between
predictions and experiment is found.
The results  of Table~\ref{efgs} suggest that the previous 
affirmation could be generalized including
perovskites, since the agreement of PCM with FLAPW 
begins to worsen when $d_{nn}(Cd-O)$ approaches this value.

In their PAC study of \cdti\ as a function of temperature,
Baumvol {et al.} obtained that \cgav\ and  $\eta$ 
increase when T decreases. This confirms the tendency predicted
in our calculations assuming a positive 
thermal expansion coefficient for \cdti.
The augment of  $\eta$ with pressure may be explained because 
the distorsion of the 
perovskite structure also increases with pressure.
This can be seen by looking at
Table~\ref{tilting} where we show the behaviour of three parameters defined
in Ref.~\onlinecite{sasaki} to caracterize the distorsion
of this and other perovskites.
 For a cubic perovskite $t_{obs}=1$, $\phi_{ab}=180^o$
and $\phi_{bc}=90^o$, and when larger is the separation from these values,
larger is the distorsion of the structure and the tilting of the octahedra.
We have obtained that the mentioned parameters vary almost linearly for
an isotropic change of volume of \cdti\ perovskite.

\begin{table} 
\caption{\label{tilting}
Parameters showing the distorsion in \cdti\ perovskite
as a function of an isotropic
change of volume.
The observed tolerance factor ($t_{obs}$) and the tilting angles of the octahedra
in ab-plane ($\phi_{ab}$) and bc-plane ($\phi_{bc}$) as defined in
 Ref.~\onlinecite{sasaki}.
$t_{obs}=\langle Cd-O \rangle /(\sqrt{2} \langle Ti-O \rangle )$, 
where $\langle Cd-O \rangle$ 
and $\langle Ti-O \rangle$ are the mean interatomic 
distances with twelve and six
coordination for Cd and Ti sites respectively.}
\begin{ruledtabular}
\begin{tabular}{lccc}
$V/V_0$ & $t_{obs}$ & $\phi_{ab}$ & $\phi_{bc}$  \\
			\hline 			  
0.913 & 0.9759  & 141.52 & 105.20  \\ 
0.956 & 0.9765  & 142.18 & 105.11   \\
1.000 & 0.9772  & 142.87 & 104.97 \\
1.046 & 0.9777  & 143.53 & 104.81 \\
1.093 & 0.9783  & 144.24 & 104.66 \\
\end{tabular}
\end{ruledtabular}
\end{table}

\begin{figure}[ht]
\includegraphics[width=9cm,height=8.5cm,angle=270]{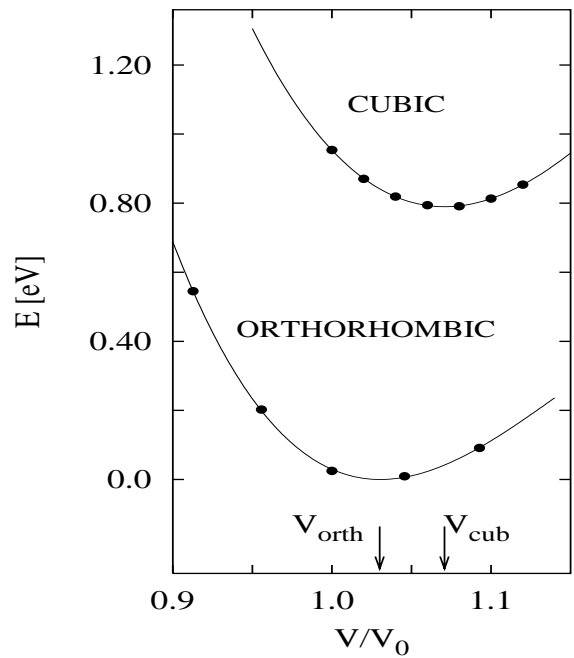}
\caption{\label{figen} Energy vs. Volume for cubic and orthorhombic
phases of CdTiO$_3$ perovskite. Energy is referred to the minimum
obtained for the orthorhombic phase, and volume is in terms of the
experimental volume of the orthorhombic phase, V$_0$
\cite{sasaki}.
Arrows indicate the positions of the minima of the plotted curves.
Full circles: calculated points, lines: cubic-fitt.}
\end{figure}

Even when in this study we keep fixed the $a/b$ and $a/c$ relations, 
and so
we don't obtain an unconstrained energy vs volume dependence, it is 
interesting to look at this E(V) 
curve compared with the one of the cubic phase (see Figure 1).
The high energy 
difference obtained between both phases (two orders of magnitude
larger than the one obtained for the same phases in BaTiO$_3$
 \cite{vanderbilt})
may explain why the transition
to cubic \cdti\ has not been reported \cite{constraint}. 
In fact, perovskites that crystallize in an orthorhombic
\pbnm\ structure at room temperature, as CaTiO$_3$, SrZrO$_3$ and SrHfO$_3$ 
undergo three transitions when the temperature is increased: 
orthorhombic ($Pbnm$) $\rightarrow$ orthorhombic ($Cmcm$)
                    $\rightarrow$ tetragonal ($I4/mcm$)
                    $\rightarrow$ cubic \cite{catio3,srzro3,srhfo3}.
For the case of
SrHfO$_3$  perovskite, for example, the transitions occur for temperatures 
around 870 K, 1000 K and 1360 K. As far as we know, no structural
study of
\cdti\ at high temperature has been performed. Moreover
in PAC studies of orthorhombic \cdti\ up to 
1270 K no signature of any phase transition has been reported \cite{catchen}.
In a previous study on SrHfO$_3$ \cite{aarhus} a
difference of 0.27 eV has been obtained between orthorhombic ($Pbnm$)
and cubic structures. Thus looking at the 0.8 eV obtained
in the present work for \cdti\ we would expect the transition
to the cubic phase, if present, to occur at very high temperature.
Future experimental work on \cdti\ at high temperatures would be valuable
to check this prediction.

In summary our calculations help to decide between the proposed structures
for orthorhombic \cdti\ perosvskite. We have obtained that within
the precision of our FLAPW calculations
the proposed
\pbnm\ structure from Ref.~\onlinecite{sasaki} is stable while
\pbnv\ is not.
We have obtained very good agreement between
the calculated EFG and experiment and have shown that PCM approximation
works quite well in this case. 
However, care should be taken when
applying PCM to other systems with lower Cd-O bond, since the agreement
seems to worsen in this direction.
Our energy calculations suggest that the transition
to the cubic phase, if present, would take place at quite high temperature. 

\begin{acknowledgments}
This work was supported by CONICET and {\em Fundaci\'on Antorchas}, 
Argentina. We want to aknowledge Dra. P. de la Presa for fruitful
discussions.
\end{acknowledgments}


\begin{thebibliography}{99}
\bibitem{pbn21}  In fact in Ref~\onlinecite{kaymiles} the space group is
                 reported as $Pc2_1n$, but we use here \pbnv\ as in
		  Ref~\onlinecite{sasaki}.
\bibitem{constraint} Note that this argument is not affected 
by the use of a constrained energy since it is an upper bound,
so the E$_{\rm cub}$ -E$_{\rm orth}$ energy difference 
is a lower bound.
\end{thebibliography}
\end{document}